\documentclass[12pt,preprint]{aastex}
\usepackage{graphics}

\begin{document}
\def\gsim{\;\rlap{\lower 2.5pt
\hbox{$\sim$}}\raise 1.5pt\hbox{$>$}\;}
\def\lsim{\;\rlap{\lower 2.5pt
   \hbox{$\sim$}}\raise 1.5pt\hbox{$<$}\;}
\def\msun{{\rm\,M_\odot}}
\newcommand\degd{\ifmmode^{\circ}\!\!\!.\,\else$^{\circ}\!\!\!.\,$\fi}
\newcommand{\etal}{{\it et al.\ }}
\newcommand{\uv}{(u,v)}
\newcommand{\rdm}{{\rm\ rad\ m^{-2}}}
\newcommand{\muas}{{\rm\ \mu as}}
\def\earth{{\oplus}}
\def\ripl{RIPL\, }

\title{Radio Astrometric Detection and Characterization of Extra-Solar Planets:
\\
A White Paper Submitted to the NSF ExoPlanet Task Force}

\author{Geoffrey C. Bower\altaffilmark{1}, Alberto Bolatto\altaffilmark{1}, Eric Ford\altaffilmark{2}, Paul Kalas\altaffilmark{1}, Jim
Ulvestad\altaffilmark{3}}

\altaffiltext{1}{Astronomy Department \& Radio Astronomy Laboratory,
University of California, Berkeley, CA 94720; gbower,bolatto,pkalas@astro.berkeley.edu}
\altaffiltext{2}{Harvard-Smithsonian Center for Astrophysics, 60 Garden St., 
Cambridge, MA  02138; ericbford@gmail.com}
\altaffiltext{3}{National Radio Astronomy Observatory, P.O. Box 0, Socorro NM
87801, U.S.A. ; julvesta@nrao.edu}

\begin{abstract}

The extraordinary astrometric accuracy of radio interferometry 
creates an important and unique opportunity for the discovery and 
characterization of exo-planets.  Currently, the Very Long Baseline Array
can routinely achieve better than 100 $\muas$ accuracy, and
can approach 10 $\muas$ with careful calibration.  
We describe here RIPL, the Radio Interferometric PLanet
search, a new program with the VLBA and the Green Bank 100 m telescope
that will survey 29 low-mass,
active stars over 3 years with sub-Jovian planet mass sensitivity at 1 AU.
An upgrade of the VLBA bandwidth will increase astrometric accuracy
by an order of magnitude.  Ultimately, the colossal collecting area of the 
Square Kilometer Array could push astrometric accuracy to 1 microarcsecond,
making detection and characterizaiton of Earth mass planets possible.

RIPL and other future radio astrometric planet searches occupy a unique
volume in planet discovery and characterization parameter space.
The parameter space of astrometric searches gives greater sensitivity
to planets at large radii than radial velocity searches.
For the VLBA and the expanded VLBA,
the targets of radio astrometric surveys are by necessity nearby, low-mass,
active stars, which cannot be studied efficiently through
the radial velocity method, coronagraphy, or optical interferometry.  
For the SKA, detection sensitivity will extend to solar-type stars.
Planets discovered
through radio astrometric methods will be suitable for characterization
through extreme adaptive optics.

The complementarity of radio astrometric techniques with other 
methods demonstrates that radio astrometry can play an important role 
in the roadmap for exoplanet discovery and characterization.  

\end{abstract}

\section{Radio Astrometry and Extra-Solar Planets}

Radio astrometry has long been the gold standard for definition of 
celestial reference frames (Fey et al. 2004, AJ 127, 3587) and has been 
used to obtain the most accurate geometric measurements of any 
astronomical technique.  Astrometric results include measurement of 
the deflection of background sources due to the gravitational
fields of the Sun and Jupiter (Fomalont \& Kopeikin 2003, ApJ, 598, 704), 
the parallax and proper motion of pulsars at distances greater than 
1 kpc (Chatterjee et al. 2005, ApJ, 630, L61), an upper limit
to the proper motion of Sagittarius A* of a few ${\rm\ km\ s^{-1}}$
(Reid \& Brunthaler 2004, ApJ, 616, 872), the rotation of the disk 
of M33 (Brunthaler et al. 2005, Science, 307, 1440), and
a $<1\%$ distance to the Taurus star-forming
cluster (Loinard 2006, BAAS, 209, 1080).

The Very Long Baseline Array (VLBA) images nonthermal radio emission
and can routinely achieve 100 $\muas$ astrometric accuracy, but has
achieved an accuracy as high as 8 $\muas$ under favorable
circumstances (Fomalont \& Kopeikin 2003).
Nonthermal stellar radio emission has been detected from many stellar
types (G\"udel 2002, ARA\&A, 40, 217), including brown dwarfs 
(Berger et al. 2001, Nat, 410, 338),
proto-stars (Bower et al. 2003, ApJ, 598, 1140) , massive
stars with winds (Dougherty et al. 2005, ApJ, 623, 447), 
and late-type stars 
(Berger et al. 2006, ApJ, 648, 629).
Only late-type
stars are sufficiently bright, numerous, nearby, and low mass to
provide a large sample of stars suitable for large-scale astrometric
exoplanet searches.  Radio astrometric searches can determine whether
or not M dwarfs, the {\em largest stellar constituent of the Galaxy}, are
surrounded by planetary systems as frequently as FGK stars and if the
planet mass-period relation varies with stellar type.  The population
of gas giants at a few AU around low mass stars is an important
discriminant between planet formation models.

Radio astrometric searches have a number of unique qualities:
\begin{itemize}

\item Opportunity to discover planets around nearby active M dwarfs at
large radii;
\item Ability to fully characterize orbits of detected planets,
without degeneracies in mass, inclination, and longitude of ascending
node;
\item Sensitivity to long-period planets with sub-Jovian masses
currently and Earth masses ultimately;
\item Complementary with existing planet searching techniques: most
targets cannot be explored through other methods;
\item Ability to follow-up detected planets with imaging and
spectroscopy; and,
\item Absolute astrometric positions within the radio reference frame
for stars and planets.
\end{itemize}

The quality and uniqueness of radio astrometry for planet searches are
the result of two factors:

	$\bullet$ {\bf High precision of radio astrometry:} The VLBA
	can routinely achieve 100 $\muas$ accuracy through
	relative astrometry.  This precision is
	an order of magnitude better than obtained from laser-guide
	star adaptive optics (e.g., Pravdo et al. 2005).
	Future instruments will have one to two orders of magnitude
	more accurate astrometry, comparable to the best accuracy
        achievable with the proposed SIM spacecraft.

	$\bullet$ {\bf Active stars are difficult to study in optical
	programs:} Our target stars are active M dwarfs,
	which have radio fluxes on the order of 1 mJy.  These radio stars
	are difficult to study through optical radial velocity
	techniques because they are faint and because the activity in
	these stars distorts line profiles, reducing the accuracy of
	radial velocity measurements.

\begin{figure}[tb]
\includegraphics[width=0.75\textwidth]{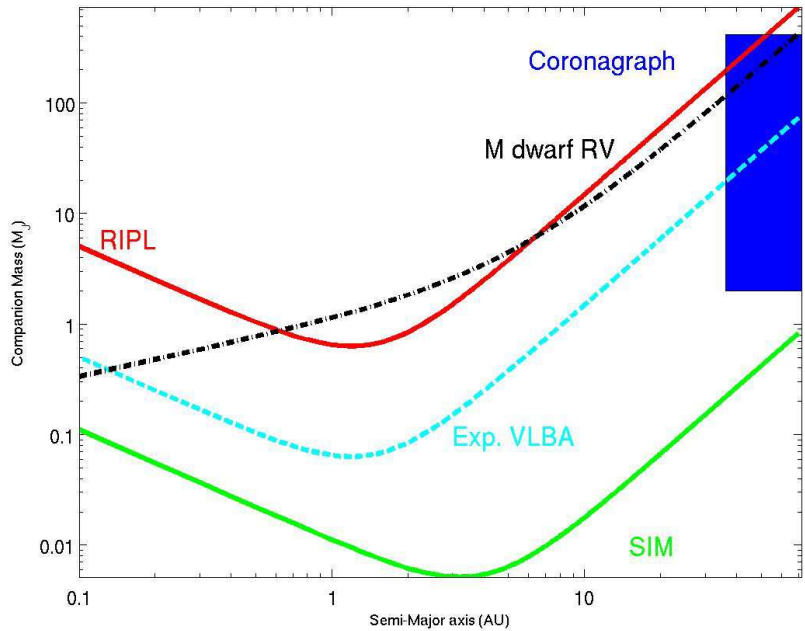}
\caption{Sensitivity of different methods in planet mass and semi-major axis
space for radio astrometric surveys and other methods.
``Exp. VLBA'' refers to the upgraded VLBA described in \S~\ref{sec:VLBA}.
The semi-major axis at the 
minimum in the astrometric search curves is determined by the search duration,
which is 3 years for RIPL and the Exp. VLBA campaign.}
\label{fig:pspace}
\end{figure}

We give a sketch of the parameter space for RIPL, future radio
astrometric searches, the Space Interferometric Mission, radial
velocity searches, and coronagraphic searches in
Figure~\ref{fig:pspace}.  A comparison of the radial velocity and
astrometric amplitudes indicates that astrometric techniques are
favored over radial velocity techniques for long period ($\gsim 1$
year) planets for these faint objects, for an astrometric accuracy of
$\sim 100 \muas$ (Ford 2006, PASP, 118, 364).

In Section 2, we describe the sensitivity and methods of radio astrometry.
In Section 3, we describe a new program with the VLBA and the Green Bank 100m
telescope to search for planets around nearby M dwarfs.  In Section 4, we
demonstrate that a bandwidth upgrade for the VLBA will increase astrometric
accuracy or stellar sample sizes by an order of magnitude.  In Section 5,
we discuss the role that the Square Kilometer Array can play with its three
order of magnitude increase in sensitivity over the VLBA.

\section{Radio Astrometry Sensitivity and Methods}

Astrometric exoplanet searches must be able to detect an astrometric
 signal that has an amplitude of
\begin{equation}
\theta = 2 {a \over D} * {M_{p} \over M_{*} } = 
1400 \muas * {a \over 1 AU} * {5 {\rm\ pc} \over D} * {M_p / M_J} * {0.2 M_\sun \over M_*},
\end{equation}
for a planet of mass $M_p$ orbiting a star of mass $M_*$ with a
semimajor axis $a$ at a distance $D$ from the Sun (a mass of 0.2
$M_\sun$ corresponds to a M5 dwarf).  To robustly detect a planet,
observations must span at least a significant fraction of a period
\begin{equation}
T = 2.2\,yr * {\left(a \over 1 AU\right)}^{3/2} * {\left(0.2 M_\sun \over M_*\right)}^{1/2}.
\end{equation}

The ultimate accuracy that can be obtained through
a radio astrometric technique is
\begin{equation}
\sigma_{ast}= \sigma_{beam} / {\rm SNR},
\end{equation}
where $\sigma_{beam}=b/\lambda$ is the synthetic beam size for an
array with maximum baseline $b$, $\lambda$ is the observing
wavelength, and SNR is the signal to noise ratio of the target source
detection. For the VLBA $\sigma_{ast}\approx500 \muas/{\rm SNR}$.

The astrometric position is defined relative to nearby ($\sim 1^\circ$)
compact radio sources.  Typical observations include switching on
minute timescales between the calibrator and the target sources,
with less frequent observations of secondary calibrators.
The use of multiple calibrators is intended to determine the differential
delay in position on the sky due to varying path length from tropospheric
water vapor.  The extent to which this cannot be calibrated sets the 
final astrometric accuracy in observations that are not SNR-limited.  
The nearer the
calibrators and the greater sensitivity at which they can be detected typically
determines this error.  The error decreases linearly with decreasing 
calibrator-target separation.  The increased sensitivity of future arrays will 
increase the calibrator density and therefore decrease the typical separation
from calibrator to target and the uncalibrated astrometric error.
For sufficiently small target to calibrator separation, the calibrator will
be in the primary beam of the antenna, enabling simultaneous
observations of the target and calibrator that also remove temporal dependence
of tropospheric variations.

\section{RIPL:  Radio Interferometric Planet Search}

RIPL is a 1400-hour, 3-year VLBA and GBT program to search
for planets around 29 nearby, low-mass, active stars.  The program will 
achieve sub-Jovian planet mass sensitivity.  The observing program will 
be completed in 2009.

The most serious limitation to astrometric accuracy may be from
stellar activity that jitters the apparent stellar position.  Most
evidence, however, indicates that this radio astrometric 
jitter is small.  For instance,
White, Lim and Kundu (1994, ApJ, 422, 293) 
model the radio emission from dMe stars as originating
within $\sim 1$ stellar radius of the photosphere.  At a distance of
10 pc for a M5 dwarf a stellar radius is $\sim 100$ $\mu$as, roughly
an order of magnitude smaller than the astrometric signature of a
Jupiter analog.  We conducted the VLBA Precursor Astrometric Survey
(VPAS) in Spring 2006 to assess the effect of stellar jitter on
astrometric accuracy (Bower et al.  2007, in prep.).

\begin{figure}[tb]
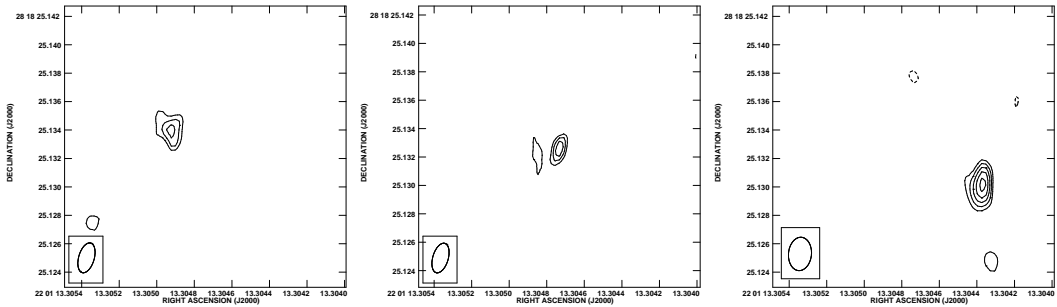

\center\mbox{\includegraphics[width=0.25\textwidth,angle=-90]{GJ4247_B2.PS}\includegraphics[width=0.25\textwidth,angle=-90]{GJ4247_C2.PS}\includegraphics[width=0.25\textwidth,angle=-90]{GJ4247_A2.PS}}
\caption[]{Images of GJ4247 in three separate epochs on 23, 25, and 
26 March 2006 (right to left) from the VLBA Precursor Astrometric Survey.  
Contour levels are -3, 3, 4, 5, 6, 7, 8 times the rms noise of 95 $\mu$Jy.  
The synthesized beam is shown in the lower left hand corner of each image.
\label{fig:motion}}
\end{figure}

For each star, three VLBA epochs were spread over fewer than 10 days.
Seven stars were detected in at least one epoch and four were detected
in all three epochs (Figure~\ref{fig:motion}).
All stars have proper motions and parallaxes determined by Hipparcos
or other optical methods with a precision of a few mas per year,
yielding predicted relative positions accurate to $\sim 100 \muas$ during
the length of the study.  For all stars detected with multiple epochs,
the motions match the results of Hipparcos astrometry well with rms in
each coordinate ranging from 0.08 to $0.26 \muas$.
Deviations in the positions appear to be limited by our
sensitivity; i.e., the effect of stellar
activity on their positions is unimportant.

\begin{figure}[tb]
\includegraphics[width=\textwidth]{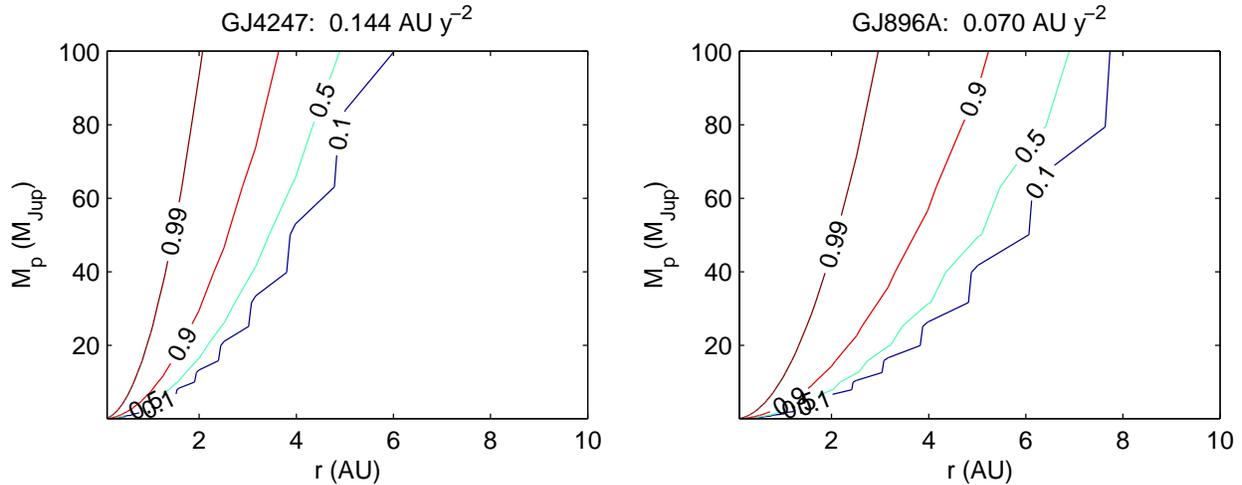}
\caption[]{Region of planetary mass and semi-major axis phase-space rejected by 
acceleration upper limits based on combination of 3 epochs of radio astrometric 
measurements and optical astrometry, primarily from Hipparcos.  Different 
contours indicate confidence intervals for excluded regions.
\label{fig:accel}}
\end{figure}

{\em In fact, the {\bf small} differences in the fitted proper motion
and the Hipparcos proper motion already eliminate brown dwarfs as
companions to these objects (Figure~\ref{fig:accel}).}  The measured
differences are consistent with noise in the VLBA astrometry ($200 \muas
{\rm\ / 3 day} \sim 20$ mas/yr).  The typical reflex motion due to
a long period brown dwarf is $\sim 100$ mas/yr, which would be
apparent.  The much longer time baseline and better sensitivity of
RIPL will reduce proper motion errors by $\sim 2$ orders of magnitude.

\subsection{Synergy with other Planet Searches}

{\em \ripl is synergistic with the existing and future planet-search 
programs, as well as current ground-based planet searches (including 
radial velocities, transits, adaptive optics, and interferometry).} 
\ripl provides an opportunity to search for planetary systems in a 
unique area of parameter space that will not be targeted by other planet 
searches until the launch of NASA SIM - Planetquest.  

Ground based transit searches are most sensitive for very short
periods ($P \sim$ days), and the Kepler mission aims to detect planets
with orbital periods of slightly more than a year.  Thus, \ripl will
make a valuable contribution to our understanding of the frequency of
long-period planets around M stars.  Further, unlike transits and
radial velocity observations astrometric measurements directly measure
the planet mass, which is important for testing models of planet
formation.  While the unknown inclination is less of an issue for
studying large samples of planets, measuring individual inclinations
will be particularly valuable for planets around M dwarfs, since a
relatively modest number of M dwarfs are being surveyed by RIPL ($\sim
30$ vs $\sim 3000$ stars by radial velocities).

Ground-based optical and near-infared interferometers (e.g., PTI,
NPOI) require bright stars and are not appropriate for faint low-mass
stars.  The \ripl astrometric accuracy is an order of magnitude
better than the astrometric error from Keck Laser Guide Star Adaptive
Optics astrometry (Pravdo et al.  2005, ApJ, 630, 528).  
{\em Thus, \ripl is the best
means for an astrometric search of M dwarfs until SIM launches} (now
estimated for no earlier than 2016).

A long-period planet detected by \ripl would enable exciting
scientific investigations such as photometric and spectroscopic
observations to determine the planets physical properties.  While
space based missions such as TPF-C and TPF-I are expected to be
extremely powerful and aim to directly detect terrestrial mass
planets, these missions are not expected to launch for at least a
decade in the future.  Knowing which stars have giant planets suitable
for direct imaging would enable direct probes of an extrasolar planet.

\section{VLBA Upgrade and Planet Detection} 
\label{sec:VLBA}

The VLBA is presently being upgraded from a typical data rate of 256
Mbit/s to 4 Gbit/s, with project completion estimated by 2010.  This
will result in a sensitivity increase by a factor of 4, or about a
factor of 8 increase in areal density of reference sources on the sky.
Thus, the typical distance between a target star and its nearest
reference source will decrease by a factor of $\sim 3$.  A few years
later we expect a data rate of 16 Gbit/s, yielding a target-calibrator
separation more than 10 times smaller than current values.  Since in
the limit of infinite SNR the astrometric error depends linearly on
the separation from the reference source, relative astrometric errors
of $\lesssim 10$~$\mu$as should be fairly routine; in principle, this
would permit detection of a planet with a mass of less than 10\% of
the mass of Jupiter. The sensitivity increase afforded by these
upgrades will also permit a sizable increase of the late-type dwarf
sample.

\section{Square Kilometer Array} 

The Square Kilometer Array (SKA; Carilli \& Rawlings 2004, New AR, 48,
979) is a proposed future radio telescope
that would have a collecting area of a square kilometer, approximately
200 times the collecting area of the VLBA.  The SKA would be built
toward the end of the next decade; it is planned to cover the
frequency range from 0.1 to 25~GHz, with the 5--10~GHz range being
most useful for astrometric planet detection.  If 25\% of the SKA area
at $\sim 8$~GHz is constructed on baselines of 1000-5000~km, it will
supply revolutionary astrometric accuracy (Fomalont \& Reid 2004, New
AR, 48, 1473).  With dish antennas of 12m diameter, the combination of
sensitivity and wide field of view often will enable many astrometric
reference sources to be found in the same antenna field of view as the
target star, allowing all temporal variations in Earth's atmosphere to
be removed.  In such a case, the relative astrometric accuracy may
reach $\sim 1$~$\mu$as, competitive with SIM and enabling astrometric
detection of Earth-mass planets.

The sensitivity of the SKA will enable astrometric detection of 
thermal emission from stars.  
The Sun, for instance, would be detectable to a distance of 10 pc with
the SKA.  Thus, the SKA will be capable of detecting and characterizing
planets around Sun-like stars.

\end{document}